\documentclass{article}

\usepackage{microtype}
\usepackage{graphicx}
\usepackage{subfigure}
\usepackage{booktabs} 

\usepackage{amsmath,amsfonts,bm}

\def\Figref#1{Figure~\ref{#1}}

\def\eqref#1{equation~\ref{#1}}

\def\1{\bm{1}}

\def\va{{\bm{a}}}

\def\ve{{\bm{e}}}

\def\vu{{\bm{u}}}
\def\vv{{\bm{v}}}

\def\vx{{\bm{x}}}

\def\evu{{u}}
\def\evv{{v}}

\def\evx{{x}}

\def\mB{{\bm{B}}}

\def\mM{{\bm{M}}}

\def\mW{{\bm{W}}}

\DeclareMathAlphabet{\mathsfit}{\encodingdefault}{\sfdefault}{m}{sl}
\SetMathAlphabet{\mathsfit}{bold}{\encodingdefault}{\sfdefault}{bx}{n}

\newcommand{\E}{\mathbb{E}}

\newcommand{\R}{\mathbb{R}}

\usepackage[accepted]{icml2024}

\usepackage{amsmath}
\usepackage{amssymb}
\usepackage{mathtools}
\usepackage{amsthm}

\usepackage[capitalize,noabbrev]{cleveref}

\theoremstyle{plain}
\newtheorem{theorem}{Theorem}[section]
\newtheorem{proposition}[theorem]{Proposition}

\theoremstyle{definition}
\newtheorem{definition}[theorem]{Definition}

\theoremstyle{remark}

\usepackage[textsize=tiny]{todonotes}

\icmltitlerunning{Emergence of Grid-like Representations by Training Recurrent Networks with Conformal Normalization}

\begin{document}

\twocolumn[
\icmltitle{Emergence of Grid-like Representations by Training Recurrent Networks with Conformal Normalization}



\icmlsetsymbol{equal}{*}

\begin{icmlauthorlist}
\icmlauthor{Dehong Xu}{yyy}
\icmlauthor{Ruiqi Gao}{yyy}
\icmlauthor{Wen-Hao Zhang}{xxx}
\icmlauthor{Xue-Xin Wei}{zzz}
\icmlauthor{Ying Nian Wu}{yyy}
\end{icmlauthorlist}

\icmlaffiliation{yyy}{Department of Statistics, University of California, Los Angeles}
\icmlaffiliation{xxx}{Lyda Hill Department of Bioinformatics and O’Donell Brain Institute, UT Southwestern Medical Center}
\icmlaffiliation{zzz}{Departments of Neuroscience and Psychology, UT Austin}

\icmlcorrespondingauthor{Dehong Xu}{xudehong1996@ucla.edu}
\icmlcorrespondingauthor{Ruiqi Gao}{ruiqigao@ucla.edu}
\icmlcorrespondingauthor{Ying Nian Wu}{ywu@stat.ucla.edu}

\icmlkeywords{Machine Learning, ICML}

\vskip 0.3in
]



\printAffiliationsAndNotice{}  

\begin{abstract}
Grid cells in the entorhinal cortex of mammalian brains exhibit striking hexagon grid firing patterns in their response maps as the animal (e.g., a rat) navigates in a 2D open environment. In this paper, we study the emergence of the hexagon grid patterns of grid cells based on a general recurrent neural network (RNN) model that captures the navigation process. The responses of grid cells collectively form a high dimensional vector, representing the 2D self-position of the agent. As the agent moves, the vector is transformed by an RNN that takes the velocity of the agent as input. We propose a simple yet general conformal normalization of the input velocity of the RNN, so that the local displacement of the position vector in the high-dimensional neural space is proportional to the local displacement of the agent in the 2D physical space, regardless of the direction of the input velocity. We apply this mechanism to both a linear RNN and nonlinear RNNs. Theoretically, we provide an understanding that explains the connection between conformal normalization and the emergence of hexagon grid patterns. Empirically, we conduct extensive experiments to verify that conformal normalization is crucial for the emergence of hexagon grid patterns, across various types of RNNs. The learned patterns share similar profiles to biological grid cells, and the topological properties of the patterns also align with our theoretical understanding. 
\end{abstract}

\section{Introduction}

The mammalian hippocampus formation encodes a ``cognitive map''~\citep{Tolman1948cognitive,o1979cog} of the animal's surrounding environment.  
In the 1970s, it was found that the rodent hippocampus contained place cells~\citep{o1971hippocampus}, which typically fired at specific locations in the environment. Several decades later, another prominent type of neurons called grid cells
\citep{hafting2005microstructure, fyhn2008grid, yartsev2011grid, killian2012map, jacobs2013direct, doeller2010evidence} were discovered in the medial entorhinal cortex.
Each grid cell fires at multiple locations that form a hexagonal periodic grid over the field \citep{Fyhn2004,hafting2005microstructure,Fuhs2006,burak2009accurate, sreenivasan2011grid, blair2007scale,Couey2013, de2009input,Pastoll2013,Agmon2020}. Grid cells interact with place cells and are believed to be involved in path integration \citep{hafting2005microstructure, fiete2008grid, mcnaughton2006path, gil2018impaired, ridler2019impaired, horner2016grid}, which calculates the agent's self-position by accumulating its self-motion, allowing the agent to determine its location even when navigating in darkness. Thus, grid cells are often considered to form an internal Global Positioning System (GPS) in the brain~\citep{Moser2016going}. While grid cells were mostly studied in the spatial domain, it was proposed that grid-like response may also exist in non-spatial and more abstract cognitive spaces~\citep{Constantinescu2016organizing,bellmund2018navigating}.

Various computational models have been proposed to explain the striking firing properties of grid cells. Traditional approach designed hand-crafted continuous attractor neural networks (CANN)~\citep{amit1992modeling,burak2009accurate,Couey2013,Pastoll2013,Agmon2020} and studied them by simulation. 
More recently two pioneering papers~\citep{cueva2018emergence,banino2018vector} learned recurrent neural networks (RNNs) on path integration tasks and demonstrated that grid patterns emerge in the learned networks. These results have been further developed in~\citep{gao2018learning,sorscher2019unified,Cueva2020, gao2021, whittington2021relating,dorrell2022actionable,xu2022conformal}. In addition to RNN models, principal component analysis (PCA)-based basis expansion models~\citep{dordek2016extracting,sorscher2019unified,stachenfeld2017hippocampus} with non-negativity constraints have been proposed to model the interaction between grid cells and place cells.

While prior work has shed much light on the grid cells, the mathematical principle and the computational mechanisms that underlie the emergence of hexagon grid patterns are still not well understood~\citep{cueva2018emergence,Sorscher2023,gao2021,Nayebi2021,Schaeffer2022}. The goal of this paper is to propose a simple and general mechanism in the recurrent neural network model of grid cells, that leads to and explains the emergence of hexagon grid patterns of grid cells. 

Specifically, to build an RNN model for grid cells, assume the activities of the population of grid cells collectively form a vector in a high-dimensional neural space. This high-dimensional vector is a representation of the 2D self-position of the agent in the 2D physical space, or in other words, a position embedding. As the agent navigates in the environment, the position embedding is transformed by a recurrent neural network that takes the velocity of the agent as input. For the recurrent network, we propose a novel conformal normalization mechanism that modulates the input velocity by the $\ell_2$-norm of the directional derivative of the transformation defined by the recurrent network. Under conformal normalization, the local displacement of the position embedding in the high-dimensional neural space is proportional to the local displacement of the agent in the 2D physical space, regardless of the direction of the input self-velocity. As a consequence, the 2D Euclidean space is embedded conformally as a 2D manifold in the neural space, and this 2D manifold forms an internal 2D coordinate system of the 2D physical environment, thus mathematically realizing the notion that grid cells form an internal GPS~\citep{Moser2016going}. We also provide a thorough theoretical understanding that draws a connection of conformal normalization to the emergence of hexagon grid patterns.

Empirically, we examine the conformal normalization mechanism for different parametrizations of recurrent networks. We start with a linear RNN that models the movement of the position embedding on the 2D manifold via matrix vector multiplication. Then we study non-linear RNNs with different activation functions, which additionally constrains the 2D manifold as the fixed points of the non-linear transformation when the input velocity is zero. Numerical experiments show that our proposed conformal normalization leads to the hexagon grid patterns in all models. The learned patterns nicely align with both theoretical predictions and empirical observations of biological grid cells. Topological analysis shows the topological properties of the learned patterns are also aligned with our theoretical understanding.  

\paragraph{Contributions.} Our work provides a novel conformal normalization mechanism of RNNs that leads to the hexagon grid patterns of grid cells. Our linear and non-linear models based on the proposed mechanism may serve as useful building blocks for future modeling of grid cells and place cells. Theoretically, we provide understanding of the connection between conformal normalization and hexagon grid patterns. Empirically, we demonstrate that conformal normalization is crucial for the emergence of hexagon grid patterns across various types of RNNs. The learned patterns are aligned with both the profile of biological grid cells and the topological properties derived from our theoretical understanding. 


\section{Background} 

In this section, we introduce the general formulation of the recurrent neural network models of grid cells, including the position embedding, recurrent transformation, and connection to place cells~\citep{gao2021}.


\subsection{Position embedding} 
	
 \begin{figure*}[h]
	\centering	
		\begin{tabular}{c|c|c}
	\includegraphics[height=.16\linewidth]{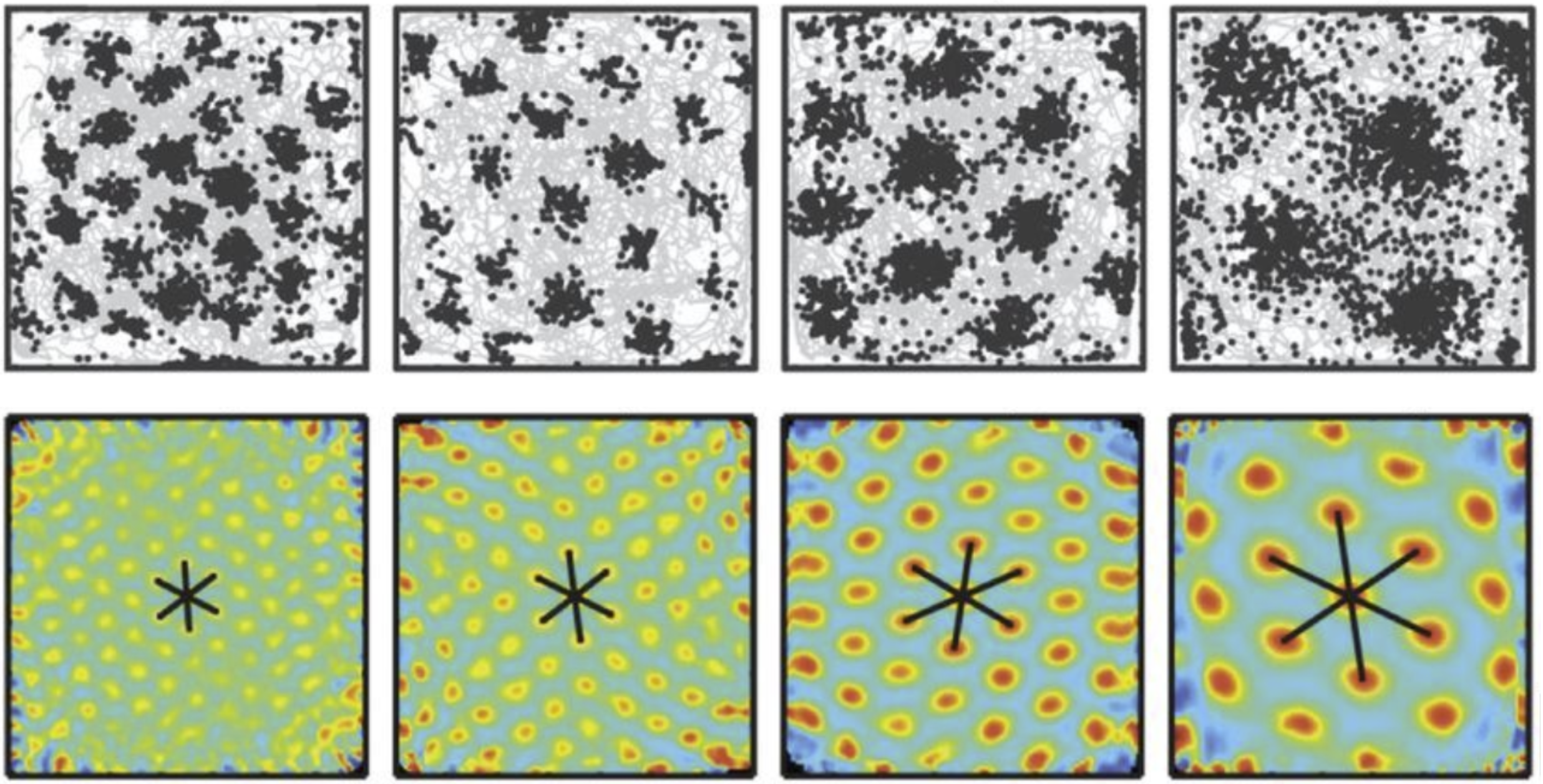} & \includegraphics[height=.18\linewidth]{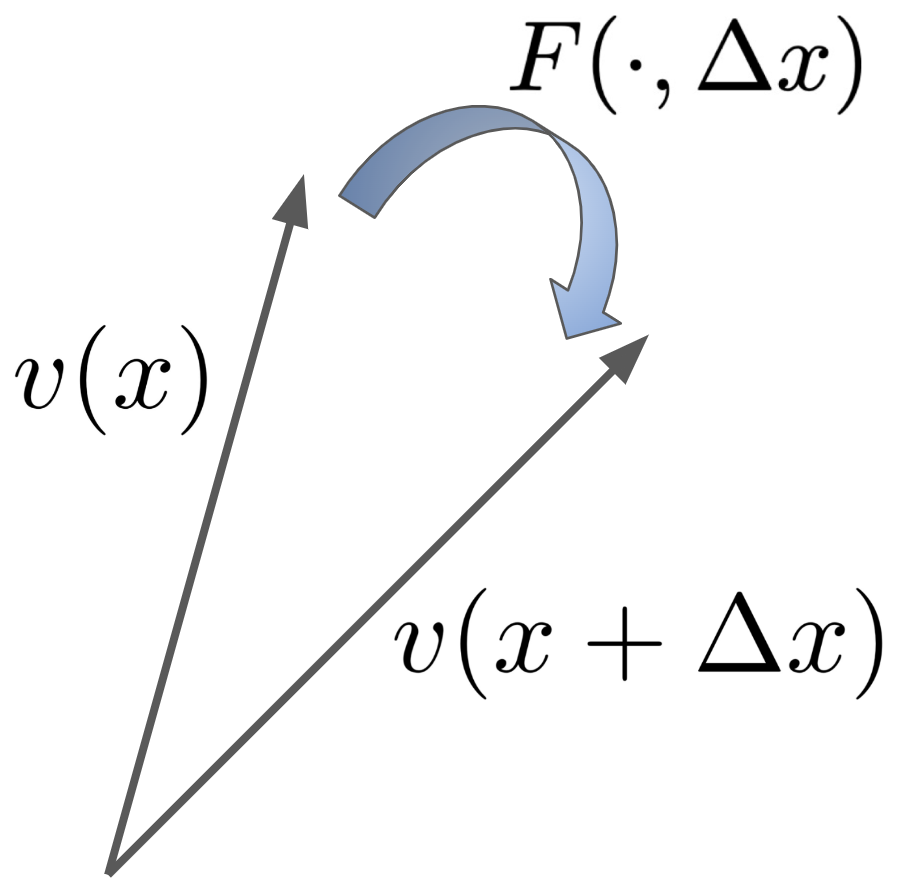} &\includegraphics[height=.18\linewidth]{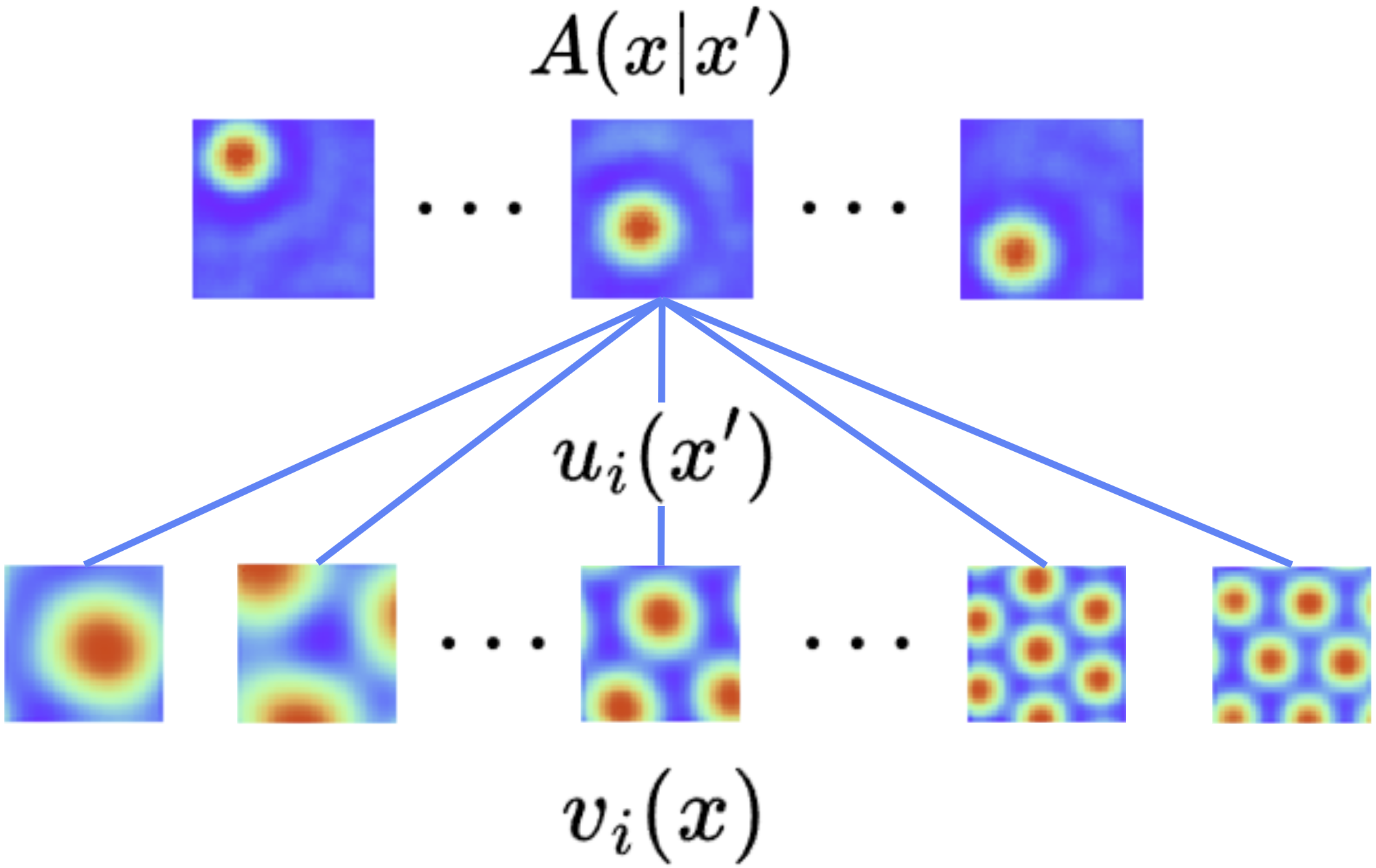}  \\
		(a) & (b) & (c) 
			\end{tabular}
		\caption{\small (a) Recorded response maps and autocorrelation maps of four different grid cells~(from \cite{moser2014network}). (b) The self-position $\vx = (\evx_1, \evx_2)$ in 2D physical space is represented by a vector $\vv(\vx)$ in the $d$-dimensional neural space. When the agent moves by $\Delta \vx$, the vector is transformed to $\vv(\vx+\Delta \vx) = F(\vv(\vx), \Delta \vx)$.  (c) Illustration of basis expansion model $A(\vx\mid \vx') = \sum_{i=1}^{d} u_i(\vx') v_i(\vx)$, where $v_i(\vx)$ is the response map of $i$-th grid cell, shown at the bottom. $A(\vx \mid \vx')$ is the response map of place cell associated with $\vx'$, shown at the top. $u_i(\vx')$ is the connection weight. All the response maps are generated by our linear model. 
  }	
	\label{fig:grid}
\end{figure*}

Suppose the agent is at the self-position $\vx = (\evx_1, \evx_2) \in \mathbb{R}^2$ within a 2D domain. The activities of the population of $d$ grid cells form a vector $\vv(\vx) = (\evv_i(\vx), i = 1, ..., d)$, where $\evv_i(\vx)$ is the activity of the $i$-th grid cell at position $\vx$. We call the space of $\vv$ the neural space, and we embed $\vx$ in the 2D physical space as a vector $\vv(\vx)$ in the $d$-dimensional neural space. It is analogous to the position embedding used by deep learning models such as Transformers~\citep{vaswani2017attention}. 


For each grid cell $i$, $\evv_i(\vx)$, as a function of $\vx$, represents the response map of grid cell $i$. The intriguing observation in neuroscience is that the response map exhibits a periodic hexagonal grid pattern, with different grid cells having varying scales, orientations, and spatial shifts (phases). Figure \ref{fig:grid}(a) displays the response maps of four different grid cells. 

\subsection{Recurrent transformation} 

At self-position $\vx = (\evx_1, \evx_2)$,  assume the agent makes a movement $\Delta \vx = (\Delta \evx_1, \Delta \evx_2)$ and moves to $\vx + \Delta \vx$. Correspondingly, the vector  $\vv(\vx)$ is transformed to $\vv(\vx+\Delta \vx)$. The general form of the transformation can be formulated as: 
\begin{eqnarray} 
    \vv(\vx + \Delta \vx)  = F(\vv(\vx), \Delta \vx),  \label{eq:PI0}
\end{eqnarray}
where $F$ can be parametrized by a recurrent neural network (RNN), and the recurrent transformation $F$ takes $\Delta \vx$ as an input. See Figure \ref{fig:grid}(b). We may call $\Delta \vx$ the input velocity if we assume a unit time period for the movement. We call (\ref{eq:PI0}) the recurrent transformation model. 

The input velocity $\Delta \vx$ can also be represented as $(\Delta r, \theta)$ in polar coordinates, where $\Delta r$ is the displacement along the direction $\theta \in [0, 2\pi]$, so that $\Delta \vx = (\Delta \evx_1 = \Delta r \cos \theta, \Delta \evx_2 = \Delta r \sin \theta)$. The transformation model  then becomes 
\begin{eqnarray} 
   \vv(\vx + \Delta \vx) = F(\vv(\vx), \Delta r, \theta),  \label{eq:PI2}
\end{eqnarray}
where we continue to use $F(\cdot)$ for the recurrent transformation. 

\subsection{Place cells} 

The vector $\vv(\vx)$ serves to inform the agent of its adjacency to different positions. This can be achieved by interaction with place cells, via a linear read-out mechanism:
\begin{eqnarray} 
    A(\vx \mid \vx') = \langle \vv(\vx), \vu(\vx')\rangle = \sum_{i = 1}^{d} \evu_i(\vx') \evv_i(\vx), \label{eq:PI1}
\end{eqnarray}
where $A(\vx \mid \vx')$, as a function of $\vx$, can be considered as the response map of the place cell associated with the place $\vx'$. In open field, the measured response map $A(\vx \mid \vx')$ can be well approximated by a Gaussian adjacency kernel $A(\vx \mid \vx') = \exp(-\|\vx - \vx'\|^2/2\sigma^2)$ for a certain scale parameter $\sigma$. $\vu(\vx') = (\evu_i(\vx'), i = 1, ..., d)$ is a $d$-dimensional read-out vector and can be regarded as the connection weight from grid cell $i$ to the place cell associated with $\vx'$. The right-hand side of Equation (\ref{eq:PI1}) implies that the response maps of grid cells $\evv_i(\vx)$ may serve as basis functions to expand the response map $A(\vx \mid \vx')$ of place cell associated with $\vx'$ (Figure \ref{fig:grid}(c)). We call (\ref{eq:PI1}) the basis expansion model. 

{\bf Path integration}. The above recurrent transformation model (\ref{eq:PI0}) and the basis expansion model (\ref{eq:PI1}) enable the agent to navigate. Suppose the agent starts from $\vx_0$, with vector representation $\vv_0 = \vv(\vx_0)$. If the agent makes a sequence of moves $(\Delta \vx_t, t = 1, ..., T)$, then the vector $\vv$ is updated by $\vv_{t} = F(\vv_{t-1}, \Delta \vx_t)$. At time $t$, the self-position of the agent can be decoded by 
$    \hat{\vx} = \arg\max_{\vx'} \langle \vv_t, \vu(\vx')\rangle$, 
i.e., the place $\vx'$ that is the most adjacent to the self-position represented by $\vv_t$. This enables the agent to infer and keep track of its position based on its self-motion even in darkness. 

\section{Conformal normalization} 

This section presents our conformal normalization mechanism for the recurrent transformation. Our mechanism is inspired by normalization in both neuroscience and deep learning. On one hand, divisive normalization is a canonical operation widely observed in the cortex and has been extensively used in previous models in computational neuroscience \citep{Geisler1992cortical, carandini2012normalization,heeger1992normalization,schwartz2001natural}, which may emerge due to the recurrent computations between the excitatory and inhibitory neurons~\citep{Rubin2015stabilized,niell2015cell}.
On the other hand, in deep learning models, batch normalization \citep{ioffe2015batch,ioffe2017batch}, layer normalization \citep{ba2016layer} and group normalization \citep{wu2018group} are ubiquitous and indispensable. 

\subsection{Definition}

Consider the general recurrent transformation
\(
    \vv(\vx+\Delta \vx) = F(\vv(\vx), \Delta r, \theta), 
\) 
where 
\(
\Delta \vx = (\Delta x_1 = \Delta r \cos \theta, \Delta x_2 = \Delta r \sin \theta),
\)
 $\theta$ is the heading direction, and $\Delta r$ is the displacement.
 
 \begin{definition}
 The \emph{directional derivative} of $F$ at $(\vv, \theta)$ is defined as 
 \begin{eqnarray}
   f(\vv, \theta) = \frac{\partial}{\partial a} F(\vv, a, \theta) \mid_{a = 0}. 
\end{eqnarray}
 \end{definition}
 With the above definition, the first order Taylor expansion of the recurrent transformation at $\Delta r = 0$ gives us 
\begin{eqnarray}
    \vv(\vx+\Delta \vx) = \vv(\vx) + f(\vv(\vx), \theta) \Delta r + o(\Delta r). \label{eq:ty_trans}
\end{eqnarray}

We define conformal normalization of the recurrent transformation so that the displacement of the position vector in the neural space is proportional to the displacement of the agent in the 2D physical space, regardless of the direction of the movement. This can be achieved by modulating the input velocity by the norm of the directional derivative of the transformation:

\begin{definition} The \emph{conformal normalization} of $\Delta r$ at $\vv(\vx)$ is defined as 
\begin{eqnarray}
   \overline{\Delta r} = \frac{s \Delta r}{\|f(\vv(\vx), \theta)\|}, \label{eq:CN0}
 \end{eqnarray}
 where $\|\cdot\|$ is the $\ell_2$ norm, and $s$ is  either a  learnable parameter or the average of $\|f(\vv(\vx), \theta)\|$ over the directions $\theta$. With conformal normalization, the transformation is changed to 
\begin{eqnarray}
    \vv(\vx+\Delta \vx) = F(\vv(\vx), \overline{\Delta r}, \theta).  \label{eq:CN1}
\end{eqnarray}
\end{definition}
\subsection{Conformal 2D manifold} 
With conformal normalization, we have a nice property of the manifold of $\vv(\vx)$ in the high-dimensional neural space. Specifically, 
\begin{proposition}
With conformal normalization (\ref{eq:CN0}) and (\ref{eq:CN1}), we have 
\begin{eqnarray}
    \|\vv(\vx+\Delta \vx) - \vv(\vx)\| = s \|\Delta \vx\| + o(\|\Delta \vx\|). \label{eq:CI}
\end{eqnarray}
 (\ref{eq:CI}) is called conformal isometry. 
\end{proposition}
The proof is straightforward. For the transformation (\ref{eq:CN1}), the first order Taylor expansion gives 
\begin{align}
\begin{split}
    \vv(\vx+\Delta \vx) &=\vv(\vx) + f(\vv(\vx), \theta) \overline{\Delta r} + o(\Delta r)\\
    &= \vv(\vx) + s \overline{f}(\vv(\vx), \theta)  \Delta r + o(\Delta r), 
\end{split}
\end{align}
where 
$
 \overline{f}(\vv, \theta) = {f(\vv, \theta)}/{\|f(\vv, \theta)\|}
$
is a unit vector with $\|\overline{f}(\vv, \theta)\| = 1$, which leads to (\ref{eq:CI}). 

Conformal isometry (\ref{eq:CI}) means that as the agent moves by $\|\Delta \vx\|$ in the 2D physical space, the position embedding $\vv$ moves by $s \|\Delta \vx\|$ in the $d$-dimensional neural space, regardless of the heading direction $\theta$. Conformal isometry leads to conformal embedding, i.e., a local coordinate system around $\vx$ (e.g., a polar coordinate system) is mapped to a local coordinate system around $\vv(\vx)$ without distortion of shape except for a scaling factor $s$. 

If $s$ is a constant over $\vx$ globally, then the 2D domain $\mathbb{D}$ is embedded as a 2D manifold $\mathbb{M} = (\vv(\vx), \vx \in \mathbb{D})$ in the neural space. This 2D manifold $\mathbb{M}$ is conformal to the 2D physical domain $\mathbb{D}$, which means $\mathbb{M}$ forms a 2D coordinate system of the physical domain without distortion except for global scaling, e.g., magnification, so that the local distance between two position embeddings informs the agent of the physical distance between the two positions. Together with the recurrent transformation $F$ that moves $\vv$ along the manifold $\mathbb{M}$, $(\mathbb{M}, F)$ effectively becomes a mathematical realization of an internal GPS system, which is widely hypothesized to be the notion of biological grid cells~\citep{Moser2016going}. 

To move one step further, the positions $\vx$ and $\vx'$ in our model do not need to be actual 2D coordinates, and there is even no need to assume an {\em a priori} 2D coordinate system. The model only needs to know the heading direction $\theta$ and self-displacement $\Delta r$ that connect different nearby positions. The nice property is, the learned $\vv$ associated with position $\vx$ will form the 2D coordinate of $\vx$ automatically. That is, our model learns to place the positions on a conformal 2D coordinate system in the high-dimensional neural space. 


\paragraph{Why high-dimensional $\vv$?} One may wonder why brains bother using a high-dimensional $\vv$ to represent a 2D coordinate. An important reason is that, a high-dimensional $\vv$ ensures to inform the agent of its adjacency to any position $\vx'$ via a linear read-out vector $\vu(\vx')$ (i.e., linear probing), even though adjacency $A(\vx| \vx')$ is highly non-linear in $\vx$ and $\vx'$. That is, $\vv(\vx)$ serves as linear basis functions to expand any non-linear value functions of $\vx$. This is related to the Peter-Weyl theory~\citep{taylor2002lectures} where group representation gives rise to linear basis functions. 

\section{Modeling}
In this section, we derive the formulation of conformal normalization for two instantiations of recurrent neural networks, namely linear and non-linear RNNs. 
\subsection{Linear model} 
 
We start by studying the following linear RNN model proposed by~\citep{gao2021}: 
\begin{align}
\begin{split}
    \vv(\vx+\Delta \vx) &= (I + \mB(\theta) \Delta r) \vv(\vx)\\
    &= \vv(\vx) + \mB(\theta) \vv(\vx) \Delta r. \label{eq:linear}
\end{split}
\end{align}
It effectively corresponds to the Taylor expansion of the transformation model at $\Delta r = 0$ (Eqn. (\ref{eq:ty_trans})), where the directional derivative is linear in $\vv(\vx)$: $f(\vv, \theta) = \mB(\theta) \vv(\vx)$. The conformal normalization is derived as
 \begin{eqnarray}
   \overline{\Delta r} = \frac{s \Delta r}{\|\mB(\theta) \vv(\vx)\|}.
 \end{eqnarray}
The linear transformation model then becomes 
\begin{eqnarray}
  \vv(\vx+\Delta \vx) =  \vv(\vx) + s\frac{ \mB(\theta) \vv(\vx) }{\|\mB(\theta)\vv(\vx)\|}\Delta r. \label{eq:linear}
\end{eqnarray}
The above model is similar to the ``add + layer norm'' operations in the Transformer model \citep{vaswani2017attention}. 

\subsection{Non-linear recurrent neural networks}
We also study the following non-linear model: 
\begin{eqnarray}
   \vv(\vx+\Delta \vx) = R(\mW \vv(\vx) + \mB(\theta) \vv(\vx) \Delta r),  \label{eq:nonlinear}
\end{eqnarray}
where $\mW$ is a learnable matrix, and $R(\cdot)$ is element-wise non-linear rectification, such as Tanh and GeLU \citep{hendrycks2016gaussian}. For this model, the directional derivative is 
\begin{eqnarray}
    f(\vv, \theta) = R'(\mW \vv) \odot \mB(\theta) \vv, 
\end{eqnarray}
where $R'(\cdot)$ is calculated element-wise, and $\odot$ is element-wise multiplication. The conformal normalization then follows (\ref{eq:CN0}) and (\ref{eq:CN1}).  

While the linear model is defined for $\vv(\vx) \in \mathbb{M}$ on the manifold, the non-linear model further constrains 
\(
   \vv(\vx) = R(\mW \vv(\vx))
\) for $\vv(\vx) \in \mathbb{M}$, where $\Delta r = 0$. If $R(\mW \vv)$ is furthermore a contraction for $\vv$ that are off $\mathbb{M}$, then $\mathbb{M}$ consists of the attractors of $R(\mW \vv)$ for $\vv$ around $\mathbb{M}$. The non-linear model (\ref{eq:nonlinear}) then becomes a continuous attractor neural network (CANN) \citep{amit1992modeling,burak2009accurate,Couey2013,Pastoll2013,Agmon2020}. See Appendix \ref{sec:more1} for more theoretical understanding of the transformation from the perspective of eigen analysis.


\subsection{Multiple blocks and multi-scale coordinate systems} 

The grid cells form multiple modules or blocks~\citep{Barry2007experience,stensola2012entorhinal}, and the response maps of grid cells within each module share the same scale. In our work, we also make a similar assumption that each module has its individual scaling factor $s$ and we hence applied conformal normalization within each module. 
As a result, each module is on a 2D manifold that serves as a coordinate system at a particular scale. With multiple modules, we have coordinate systems of multiple scales or resolutions. The notion of local distance also changes with scale. 

\subsection{Learning} 

To learn the system, we discretize $\vx \in \mathbb{D}$ and $\theta \in [0, 2\pi]$. The input consists of place cell adjacency kernels $A(\vx \mid \vx')$. The output consists of $(\vv(\vx), \vu(\vx'), \mB(\theta))$ for the linear model, and additionally $\mW$ for the non-linear model. The loss consists of two terms specifying the basis expansion model (Eqn. (\ref{eq:PI1})) and the transformation model (Eqn. (\ref{eq:PI0})) respectively: 
\begin{eqnarray}
   &L_0 = \E_{\vx, \vx'}[(A(\vx| \vx') - \langle \vv(\vx), \vu(\vx')\rangle)^2], \\
   &L_1 = \E_{\vx, \Delta \vx}[ \|\vv(\vx+\Delta \vx) - F(\vv(\vx), \Delta \vx)\|^2],
\end{eqnarray}
where $F()$ is the transformation after conformal normalization. The expectations can be approximated by Monte Carlo averages of random samples of $\vx$, $\vx'$, and $\Delta \vx$ within their ranges.  We assume $\vu(\vx') \geq 0$ because the connections from grid cells to place cells are excitatory \citep{zhang2013optogenetic,Rowland2018}. 


In the numerical experiments, we jointly learn the position embedding $\vv(\vx)$, read-out weights $\vu(\vx')$, and transformation model $F()$  by minimizing the total loss: $L = L_0 + \lambda_1 L_1$. A special case of $L_1$ with $\Delta \vx=0$ enforces the fixed point condition. 

\section{Theoretical understanding}
\label{sec: theory}
 \begin{figure}[h]
	\centering	
		\begin{tabular}{c|c|c|c}
	\includegraphics[height=.16\linewidth]{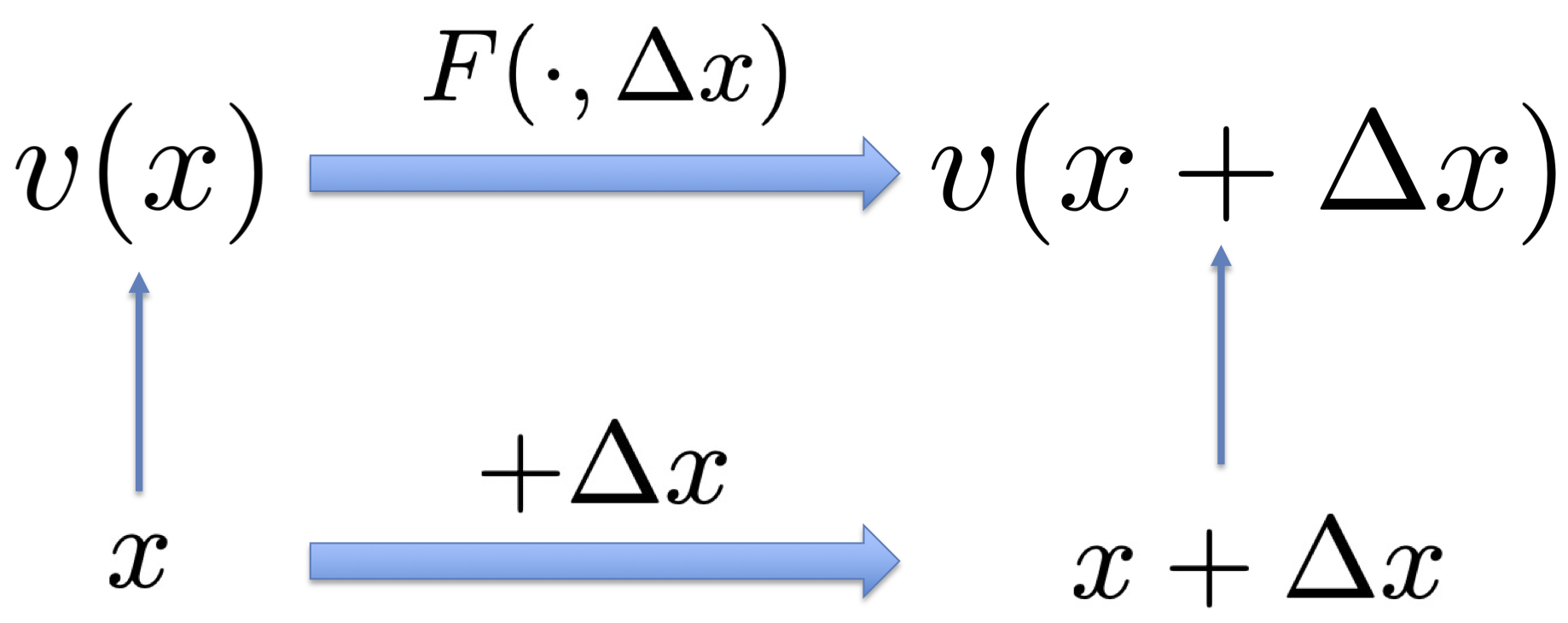}&\includegraphics[height=.15\linewidth]{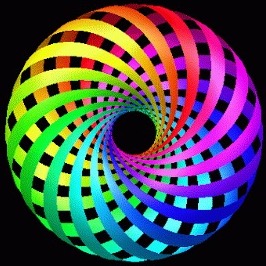}  &\includegraphics[height=.15\linewidth]{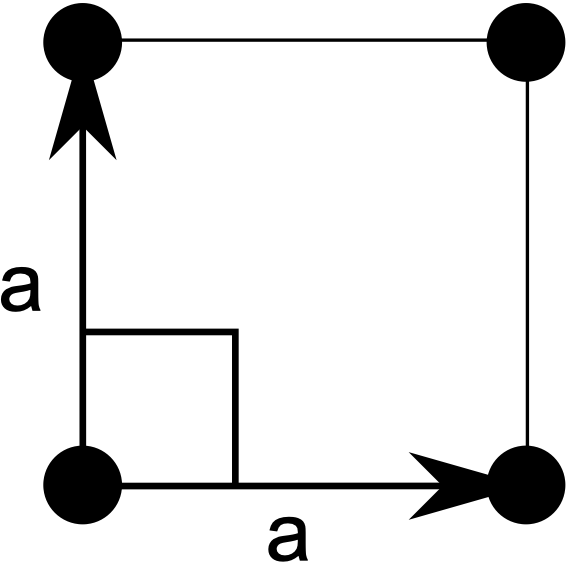} &\includegraphics[height=.15\linewidth]{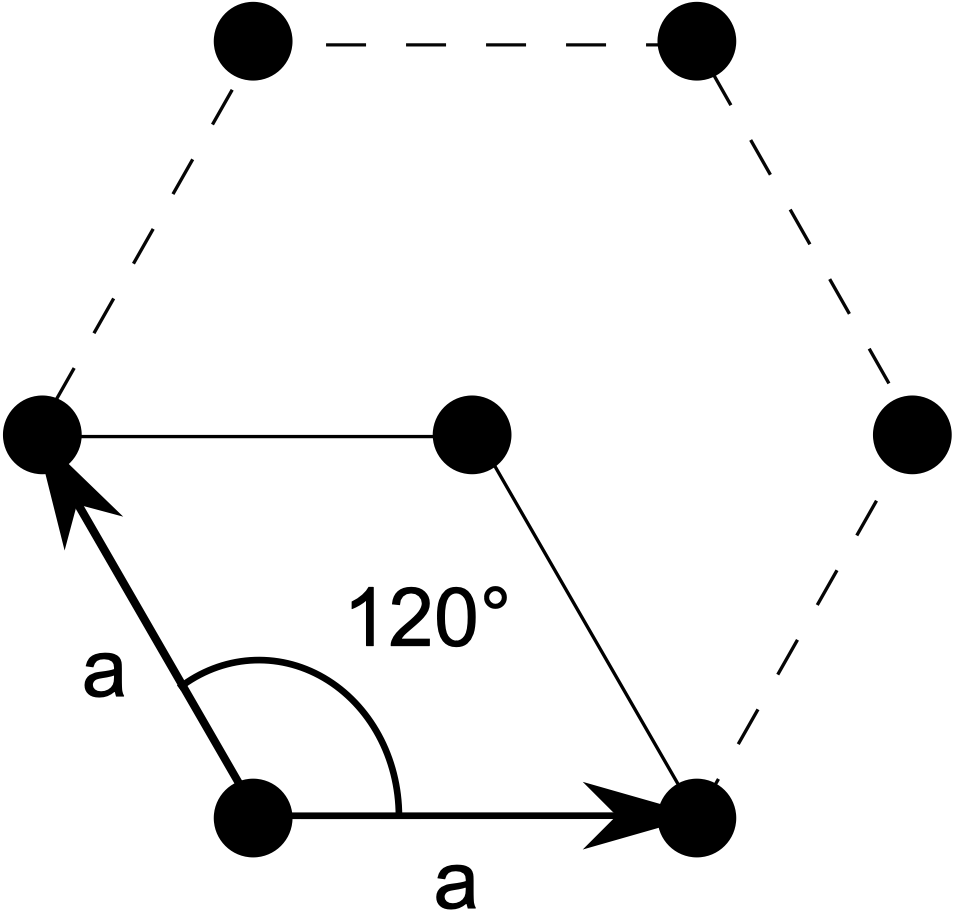} \\
		(a) & (b)  & (c) & (d)
			\end{tabular}
		\caption{\small (a) $(F(\cdot, \Delta \vx), \forall \Delta \vx \in \R^2)$ is a group of transformations, and this transformation group is a representation of the 2D additive Euclidean group $(\R^2, +)$. (b) A 2D torus embedded in 3D space~(from \cite{torus}). (c) Square lattice. (d) Hexagon lattice. }	
	\label{fig:group}
\end{figure}

In this section, we seek to connect our conformal normalization to the hexagon grid pattern in the general setting. 

\begin{figure*}[ht]
\centering
\includegraphics[width=.9\linewidth]{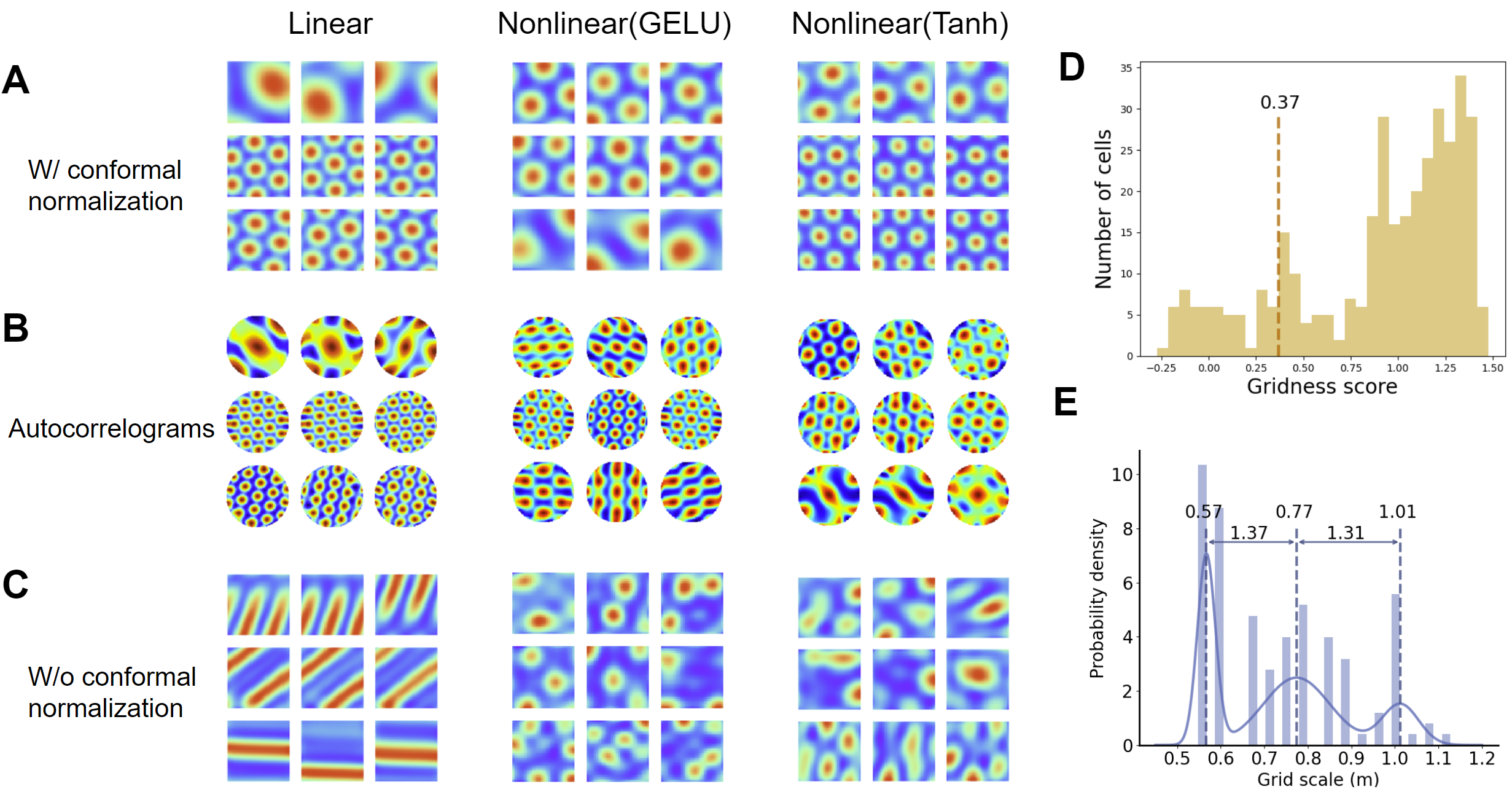} \\
  \caption{\small Results of linear and non-linear models with GELU and Tanh activation and path integration. (A) Hexagonal grid firing patterns emerge in learned \( \vv(\vx) \) across all models with conformal normalization. (B) Autocorrelograms of the learned firing patterns. (C) Without the conformal normalization condition, patterns are not hexagon grid-like. (D) Gridness score distribution of all learned grid cells. The dashed line indicates the threshold of successfulness. (E)  Multi-modal distribution of grid scales of the learned grid cells, as well as the scale ratios between successive modes. 
  }
  \label{fig: hexagon}
\end{figure*}


{\bf Step 1: Abelian Lie group}.  The group of transformation $(F(\cdot, \Delta \vx), \forall \Delta \vx)$ acting on the manifold $(\vv(\vx), \forall \vx)$ forms a representation of the 2D additive Euclidean group $(\R^2, +)$, i.e.,  $F(\vv(\vx), \Delta \vx_1 + \Delta \vx_2) = F(F(\vv(\vx), \Delta \vx_1), \Delta \vx_2) = F(F(\vv(\vx), \Delta \vx_2), \Delta \vx_1), \; \forall \vx, \Delta \vx_1, \Delta \vx_2$, and $F(\vv(\vx), 0) = \vv(\vx), \; \forall \vx$.  See Figure \ref{fig:group}(a) for an illustration. Since $(\R^2, +)$ is an abelian Lie group,   $(F(\cdot, \Delta \vx), \forall \Delta \vx)$ is also an abelian Lie group. 

{\bf Step 2: Torus topology}. Because the elements of $\vv(\vx)$ are neuron firing rates, they are bounded. Thus the manifold $(\vv(\vx), \forall \vx)$ is compact, and $(F(\cdot, {\Delta \vx}), \forall \Delta \vx)$ is a compact group. It is also connected because the 2D domain is connected. According to a classical theorem in Lie group theory \citep{dwyer1998elementary}, a compact and connected abelian Lie group has a topology of a \emph{torus}, i.e., $\mathbb{S}_1^r$, where each $\mathbb{S}_1$ is topologically a circle, and $r$ is the rank or dimensionality of the torus.

There are multiple blocks (or modules) in $\vv(\vx)$, each of which is operated separately by conformal normalization. We may assume each block (or module) has the minimal rank 2 (rank 1 can be considered a degenerate special case, see below). Otherwise, we can continue to divide the block into sub-blocks, each of which operates separately. For notation simplicity, we continue to use $F(\cdot, \Delta \vx)$ and $\vv(\vx)$ to denote the transformation and position embedding of a single block. If the torus formed by $(F(\cdot, {\Delta \vx}), \forall \Delta \vx)$ is 2D, then its topology is $\mathbb{S}_1 \times \mathbb{S}_1$, where each $\mathbb{S}_1$ is a circle.  Thus we can find two 2D vectors $\Delta \vx_1$ and $\Delta \vx_2$, so that $F(\cdot, \Delta \vx_1) = F(\cdot, \Delta \vx_2) = F(\cdot, 0)$. 
As a result, $\vv(\vx)$ is a \emph{2D periodic function} so that $\vv(\vx + k_1 \Delta \vx_1 + k_2 \Delta \vx_2) = \vv(\vx)$ for arbitrary integers $k_1$ and $k_2$. We assume $\Delta \vx_1$ and $\Delta \vx_2$ are the shortest vectors that characterize the above 2D periodicity. According to the theory of 2D Bravais lattice \citep{ashcroft1976solid} (see Appendix \ref{sec:more3} for details), any 2D periodic lattice can be defined by two primitive vectors $(\Delta \vx_1, \Delta \vx_2)$ (rank 1 degenerate case corresponds to one primitive vector being 0). 
The torus topology is supported by neuroscience data~\citep{gardner2022toroidal}.
A 2D torus is commonly visualized as a donut shape in 3D space, as shown in Figure \ref{fig:group}(b). But it can be more naturally imagined as a 2D rectangle with periodic boundary conditions. 

If the scaling factor $s$ is constant over different positions, then as the position $\vx$ of the agent moves from 0 to $\Delta \vx_1$ in the 2D space, $\vv(\vx)$ traces a perfect circle of circumference $s \|\Delta \vx_1\|$ in the neural space due to conformal isometry, i.e., the geometry of the trajectory of $\vv(\vx)$ is a perfect circle up to bending or folding but without distortion by stretching. The same with movement from 0 to $\Delta \vx_2$. Since we normalize $\|\vv(\vx)\|$ to be a constant, the two circles have the same radius and thus they also have the same circumferences, hence we have $\|\Delta \vx_1 \| = \|\Delta \vx_2\|$ (which also implies that rank 1 degenerate case is forbidden by conformal normalization). According to Bravais lattice theory \citep{ashcroft1976solid}, the periodic lattice with two equal-length primitive vectors can only be \emph{square or hexagon}, as illustrated by Figure \ref{fig:group}(c) and (d). 
  
{\bf Step 3: Fourier analysis}.  The Fourier transform of a 2D period function $f(\vx)$ can be written as a linear superposition of Fourier components $f(\vx) = \sum_k \hat{f}(\omega_k) e^{i \langle \omega_k, \vx\rangle}$, where $\omega_k = k_1 \va_1 + k_2 \va_2$,  $k = (k_1, k_2)$ are two integers, and $(\va_1, \va_2)$ are primitive vectors in the reciprocal space. For square or hexagon lattice with $\|\Delta \vx_1\|  = \|\Delta \vx_2\| = \rho$, we have $\|\va_1\| = \|\va_2\| = 2\pi/\rho$, and the lattice in the reciprocal space remains to be square or hexagon respectively. 
For a 2D Gaussian adjacent kernel centered at origin, $A(\vx) = \frac{1}{2 \pi \sigma^2}\exp(-\|\vx\|^2/2\sigma^2)$, its 2D Fourier transform is $\hat{A}(\omega) = \exp(-\sigma^2 \|\omega\|^2/2)$, which goes to zero as $\|\omega\| \rightarrow \infty$. Therefore we only need to consider frequency components $\Omega = \{\omega: \|\omega\|\leq D\}$ for a big enough $D$. For each $\vx$, let $\ve(\vx) = (e^{i \langle \omega_k, \vx\rangle}, \omega_k \in \Omega)$ be the column vector formed by the Fourier components within $\Omega$. Let $\vv(\vx) = \mM \ve(\vx)$ for a matrix $\mM$. Let us assume the dimensionality of $\vv(\vx)$ is no less than the dimensionality of $\ve(\vx)$. Then the least square regression on $\vv(\vx)$ amounts to the least squares regression on $\ve(\vx)$. The hexagon lattice packs more Fourier components into $\Omega$ than the square lattice with the same $\|\va_1\| = \|\va_2\| = 2\pi/\rho$. 
All these discrete Fourier components are orthogonal to each other. Thus the hexagon $\vv(\vx)$ provides a better least squares fit to the kernel function $A(\vx)$ in the basis expansion. Different blocks have different scales of $\|\va_1\| = \|\va_2\|$ as well as different orientations to pave the whole frequency domain. Our results expand upon previous theoretical arguments using optimal packing density to justify the optimality of hexagonal grids~\citep{Wei2015, mathis2015probable}. See Appendix \ref{sec:more} for more on theoretical understanding. 


\section{Experiments}

We conducted numerical experiments to train both linear and non-linear models. For the non-linear model, we tried two activation functions, which are GELU and Tanh, to show the generality of our theory. Within these experiments, we use a square open area measuring 1m$ \times$1m that was subdivided into $40 \times 40$ spatial bins. The dimensions of $\vv(\vx)$, representing the total number of grid cells, are 360 for the linear model and 192 for the non-linear model. 

For the response map of the place cell associated with $\vx'$, we use the Gaussian adjacency kernel with $A(\vx| \vx')=\exp(-\|\vx-\vx'\|^2/(2\sigma^2))$, where $\sigma=0.07$. For transformation, the one-step displacement $\Delta r$ is set to be smaller than $3$ grids. The scaling factor $s$ is taken to be the average of $\|f(\vv(\vx), \theta)\|$ over $\theta$. $s$ can also be a learnable parameter, and Appendix \ref{sec:pattern} contains result with learnable $s$.

\subsection{Hexagon patterns} 

Figure \ref{fig: hexagon} shows the learned firing patterns of $\vv(\vx) = (v_i(\vx), i = 1, ..., d)$  over the $40 \times 40$ lattice of $\vx$ for linear and non-linear models. In \Figref{fig: hexagon}(A), each image represents the response map for a grid cell, with every row displaying the units learned within the same module. The emergence of hexagonal patterns in these activity patterns is evident. Consistency in scale and orientation is observed within each module, though variations in phases or spatial shifts are apparent, as demonstrated in the spatial autocorrelograms of Figure \ref{fig: hexagon}(B). Our findings highlighted the essential role of conformal normalization; in its absence, as shown in Figure \ref{fig: hexagon}(C), the response maps displayed non-hexagon or stripe-like patterns. More detailed ablation results can be found in Appendix \ref{sec:ablation}. To show generality, we also provide results for both models with different block sizes.


To evaluate how closely the learned patterns align with regular hexagonal grids, we recruit the most commonly used metric for quantifying grid cells, the gridness score, adopted by the neuroscience literature~\citep{langston2010development, sargolini2006conjunctive}. We report the gridness scores and the successful rate in Table~\ref{table:gridness}, and compare with other learning-based approaches. The successful rate is computed as the percentage of grid cells with a gridness score greater than $0.37$. We showcase the grid score distribution of all learned cells in \Figref{fig: hexagon}(D). Additionally, \Figref{fig: hexagon}(E) shows the histogram of grid scales of the learned grid cell neurons, which follows a multi-modal distribution. The distribution is best fitted by a mixture of 3 Gaussians with means of $0.47$, $0.77$, and $1.01$. The ratios between successive modes are $1.37$ and $1.31$. The scale relationships in our learned grid patterns across modules align closely with both theoretical predictions~\citep{stemmler2015connecting, wei2015principle} and the empirical observations from rodent
grid cells~\citep{stensola2012entorhinal}. 

\begin{table}[h]
\caption{Gridness scores and validity rates of grid cells in learned models. The final two rows represent the results of our models. Compared to existing models, our models exhibit notably good gridness scores ($\uparrow$) and a high percentage of valid grid cells ($\uparrow$).}\label{table:gridness}
\label{sample-table}
\begin{center}
\begin{small}
\begin{sc}
{\setlength{\tabcolsep}{0.5em}
 \begin{tabular}{lcc} 
    \toprule
    Model &  Gridness score  & Valid rate \\
    \midrule
    \cite{banino2018vector} & 0.18 & 25.2\% \\
    \cite{sorscher2019unified} & 0.48 & 56.1\% \\   
    \cite{gao2021} & 0.90 & 73.1\% \\
    Ours (Linear) & {0.86} & {82.5\%} \\
    Ours (Non-linear) & {0.87} & {87.6\%} \\
    \bottomrule 
    \end{tabular}}
\end{sc}
\end{small}
\end{center}
\vskip -0.1in
\end{table}

\subsection{Path integration} 

\begin{figure}[h]
   \centering
  \includegraphics[width=.5\textwidth]
  {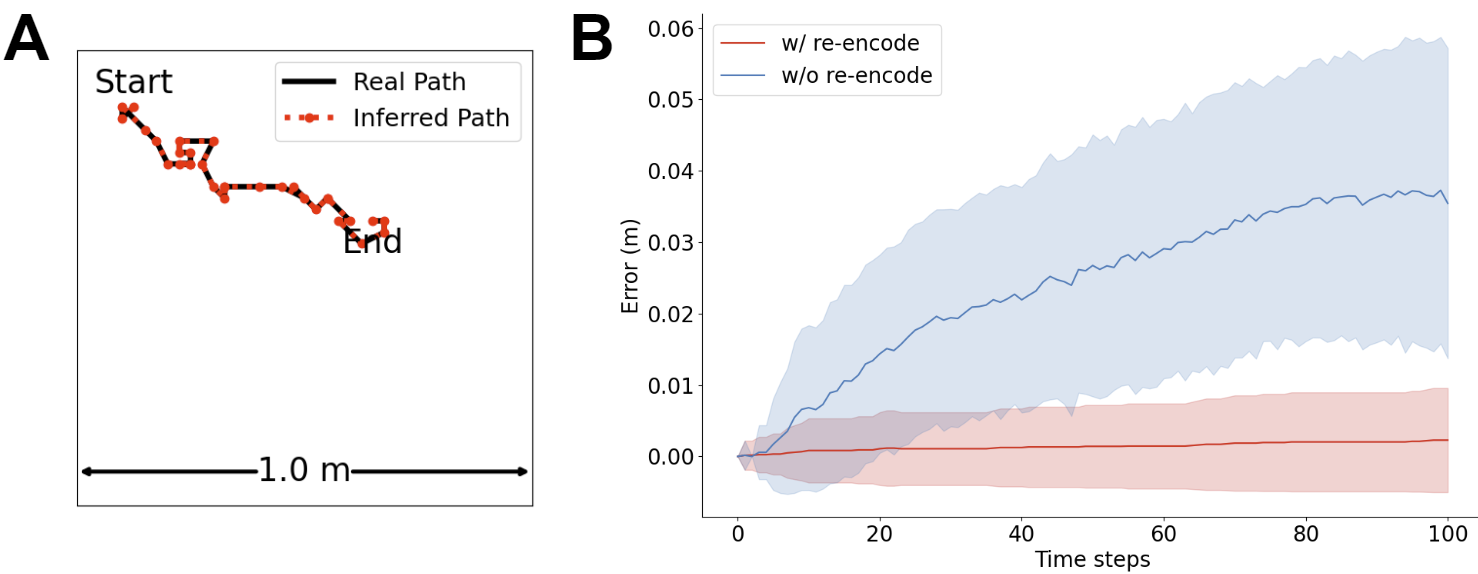}
   \caption{\small Results for path integration. (A) Path integration for 30 steps without re-encoding. The black line represents the real trajectory and the red one is the predicted trajectory by the learned model. (B) Results for long-distance (100-step) path integration error with and without re-encoding over time by the non-linear model. }
   \label{fig: path}
\end{figure}

\begin{figure*}[h]
\centering
\includegraphics[width=.8\linewidth]{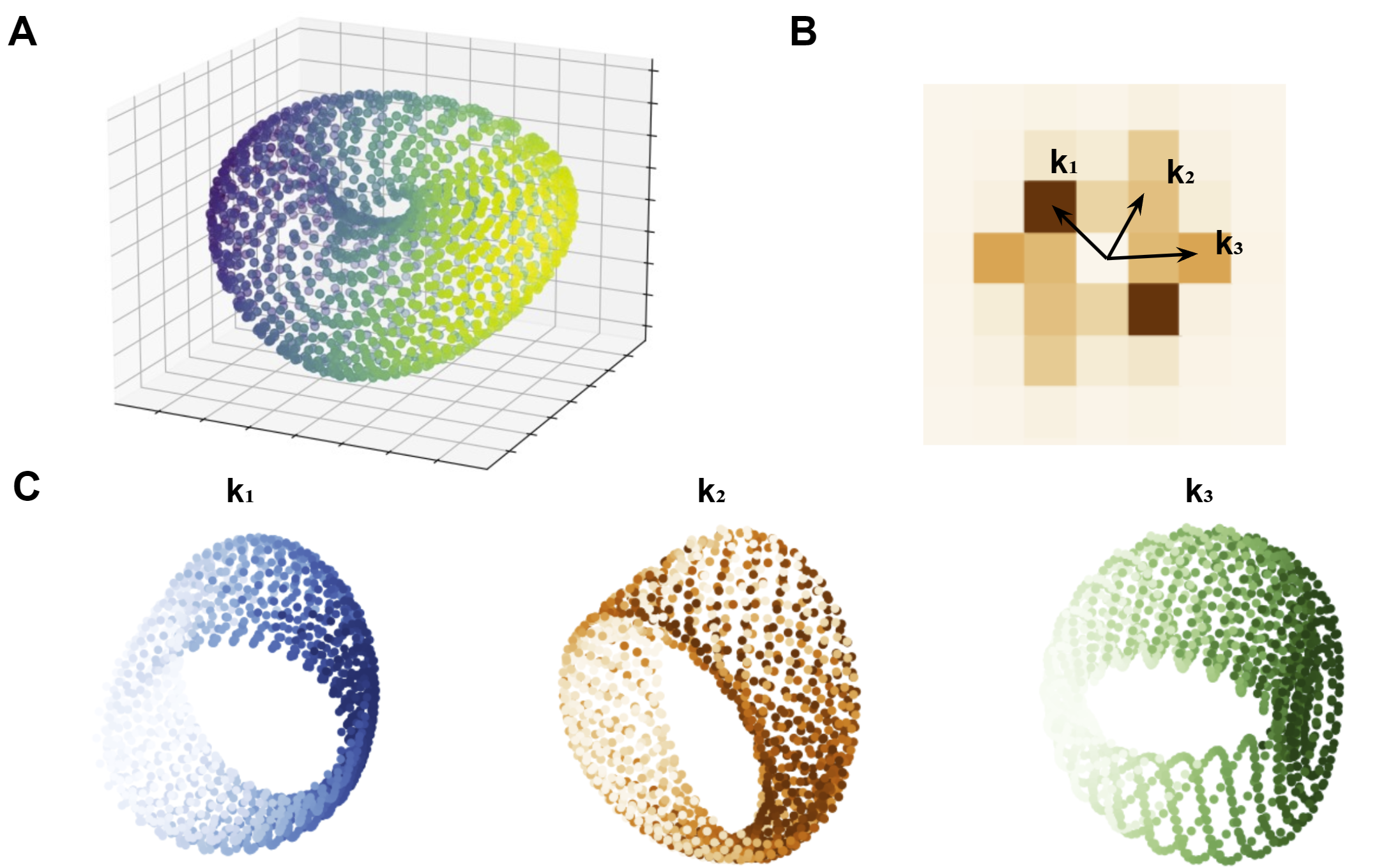} \\
  \caption{\small Toroidal structure spectral analysis in the activity of a module of grid cells. (A) Nonlinear dimensionality reduction reveals a torus-like structure in the population activity of learned grid cells. (B) Displays of the mean Fourier power spectral density, where highlighting peaks are arranged in a hexagonal pattern. (C) Projections of the reduced manifold onto three principal directional vectors, $k_1$, $k_2$, and $k_3$, which are depicted as three rings. }
  \label{fig:torus}
\end{figure*}

We assess the ability of the learned model to execute accurate path integration in two different scenarios. First of all, for path integration with re-encoding, we decode $\vv\rightarrow \hat{\vx}$ to the 2D physical space via $\hat{\vx} = \arg\max_{\vx'} \langle \vv, \vu(\vx')\rangle$, and then encode $\vv \leftarrow \vv(\hat{\vx})$ back to the neuron space intermittently. This approach aids in rectifying the errors accumulated in the neural space throughout the transformation. Conversely, in scenarios excluding re-encoding, the transformation is applied exclusively using the neuron vector $\vv$. In \Figref{fig: path}(A), the model adeptly handles path integration up to 30 steps (short distance) without the need for re-encoding. The figure illustrates trajectories with a fixed step size of three grids, enhancing the visibility of discrepancies between predicted and actual paths. It is important to note that the physical space was not discretized in our experiments, allowing the agent to choose any step size flexibly. For path integration with longer distances, we evaluate the learned model for 100 steps over 300 trajectories. As shown in \Figref{fig: path}(B), with re-encoding, the path integration error for the last step is as small as $0.003$, while the average error over the 100-step trajectory is $0.002$. Without re-encoding, the error is relatively larger, where the average error over the entire trajectory is approximately $0.024$, and it reaches $0.037$ for the last step. 


\subsection{Topological analysis} 

As we discussed in Section \ref{sec: theory}, the joint activity of grid cells from an individual module should reside on a torus-like manifold, and the positions on the torus correspond to the physical locations of a moving agent. 

To evaluate whether our empirically learned representations align with the topological properties of theoretical models, we employed a non-linear dimensionality reduction method (spectral embedding~\cite{saul2006spectral}) to show that grid cell states fell on a toroidal manifold as depicted in ~\Figref{fig:torus}(A). To further investigate periodicity and orientation within the same module, we conducted numerical simulations of pattern forming dynamics. In ~\Figref{fig:torus}(B), we applied 2D Fourier transforms of the learned maps, revealing that the Fourier power is hexagonally distributed along 3 principal directions $k_1$, $k_2$, and $k_3$. Following ~\cite{schoyen2022coherently, schaeffer2023self}, projecting the toroidal manifold onto the 3 vectors, we can observe 3 rings in ~\Figref{fig:torus}(C). This indicates the manifold has a 2D twisted torus topology. 



\section{Related work}

In computational neuroscience, hand-crafted continuous attractor neural networks (CANN)~\citep{amit1992modeling,burak2009accurate,Couey2013,Pastoll2013,Agmon2020} were designed for path integration. 
In machine learning, the pioneering papers~\citep{cueva2018emergence,banino2018vector} learned RNNs for path integration. However, RNNs do not always learn hexagon grid patterns. PAC-based basis expansion models~\citep{dordek2016extracting, stachenfeld2017hippocampus} and some theoretical accounts based on learned RNNs~\citep{Sorscher2023} rely on non-negativity assumption and the difference of Gaussian kernels for the place cells to explain the hexagon grid pattern. \citep{dorrell2022actionable} proposes an optimization-based approach to learn grid cells. 

Recent work ~\citep{schaeffer2022no} showed the prior works require hand-crafted and non-biological plausible readout representation or extra loss term~\citep{gao2021,xu2022conformal}, while, in our work, the conformal isometry is built into the recurrent network {\em intrinsically} via a simple yet general normalization mechanism so that there is no need for extra loss term. Additionally, our paper generalized the theory into non-linear models extensively and also provides a deeper and more comprehensive theoretical understanding. 



\section{Conclusion} 

Divisive normalization has been extensively studied in neuroscience and is ubiquitous in modern deep neural networks. This paper proposes a conformal normalization mechanism for recurrent neural networks of grid cells, and shows that the conformal normalization leads to the emergence of hexagon grid patterns. The proposed normalization mechanism leads to a conformal embedding of the 2D Euclidean space in the high-dimensional neural space, formalizing the notion that grid cells collectively form an internal GPS within a brain. 
\newpage
\nocite{langley00}

\bibliography{example_paper, neurips_2021,cvpr, iclr2023_conference,wureferences, pmlr-sample}
\bibliographystyle{icml2024}

\newpage
\appendix
\onecolumn
\section{Appendix}

\subsection{More theoretical understanding}
\label{sec:more}

\subsubsection{Eigen analysis of transformation}
\label{sec:more1}

For the general transformation, $\vv(\vx+\Delta \vx) = F(\vv(\vx), \Delta \vx)$, we have 
\begin{eqnarray}
  && \vv(\vx) = F(\vv(\vx), 0), \\
  && \vv(\vx+\Delta \vx) = F(\vv(\vx+\Delta \vx), 0),
\end{eqnarray}
thus 
\begin{eqnarray}
   \Delta \vv = \vv(\vx+\Delta \vx) - \vv(\vx) = F'_\vv(\vv(\vx)) \Delta \vv + o(\|\Delta \vv\|),  
\end{eqnarray}
where 
\begin{eqnarray}
   F'_\vv(\vv) = \frac{\partial}{\partial \Delta} F(\vv+\Delta, 0) \mid_{\Delta = 0}. 
\end{eqnarray}
Thus $\Delta \vv$ is in the 2D eigen subspace of $F'_\vv(\vv(\vx))$ with eigenvalue 1. 

At the same time, 
\begin{eqnarray}
   \vv(\vx+\Delta \vx) = F(\vv(\vx), \Delta \vx) = \vv(\vx) + F'_{\Delta \vx}(\vv(\vx)) \Delta \vx, 
\end{eqnarray}
where 
\begin{eqnarray}
   F'_{\Delta \vx}(\vv) = \frac{\partial}{\partial \Delta \vx} F(\vv, \Delta \vx) \mid_{\Delta \vx = 0}. 
\end{eqnarray}
Thus 
\begin{eqnarray}
   \Delta \vv = F'_{\Delta \vx}(\vv(\vx)) \Delta \vx,
\end{eqnarray}
that is, the two columns of $F'_{\Delta \vx}(\vv(\vx))$ are the two vectors that span the eigen subspace of $F'_\vv(\vv(\vx))$ with eigenvalue 1. If we further assume conformal embedding which can be enforced by conformal normalization, then the two column vectors of $F'_{\Delta \vx}(\vv(\vx))$ are orthogonal and of equal length, so that $\Delta \vv$ is conformal to $\Delta \vx$. 

The above analysis is about $\vv(\vx)$ on the manifold. We want the remaining eigenvalues of $F'_\vv(\vv(\vx))$ to be less than 1, so that, off the manifold, $F(\vv, 0)$ will bring $\vv$ closer to the manifold, i.e., the manifold consists of attractor points of $F$, and $F$ is an attractor network. 

\subsubsection{Permutation group}
\label{sec:more2}

The learned response maps of the grid cells in the same module are shifted versions of each other, i.e., there is a discrete set of displacements $\{\Delta \vx\}$, such as for each $\Delta \vx$ in this set, we have $\vv_i(\vx + \Delta \vx) = \vv_j(\vx)$, where $j = \sigma(i, \Delta \vx)$, and $\sigma$ is a mapping from $i$ to $j$ that depends on $\Delta \vx$. In other words, $\Delta \vx$ causes a permutation of the indices of the elements in $\vv(\vx)$, and $F(\cdot, \Delta \vx) \cong \sigma(\cdot, \Delta x)$, that is, the transformation group is equivalent to a subgroup of the permutation group. This is consistent with hand-designed CANN. A CANN places grid cells on a 2D ``neuron sheet'' with periodic boundary condition, i.e., a 2D torus, and lets the movement of the ``bump'' formed by the activities of grid cells mirror the movement of the agent in a conformal way, and the movement of the ``bump'' amounts to cyclic permutation of the neurons. Our model does not assume such an {\em a priori} 2D torus neuron sheet, and is much simpler and more generic.

\subsubsection{Background on Bravais lattice}
\label{sec:more3}

Named after Auguste Bravais (1811-1863), the theory of Bravais lattice was developed for the study of crystallography in solid state physics. 

\begin{figure}[h]
	\centering	
	\includegraphics[height=.15\linewidth]{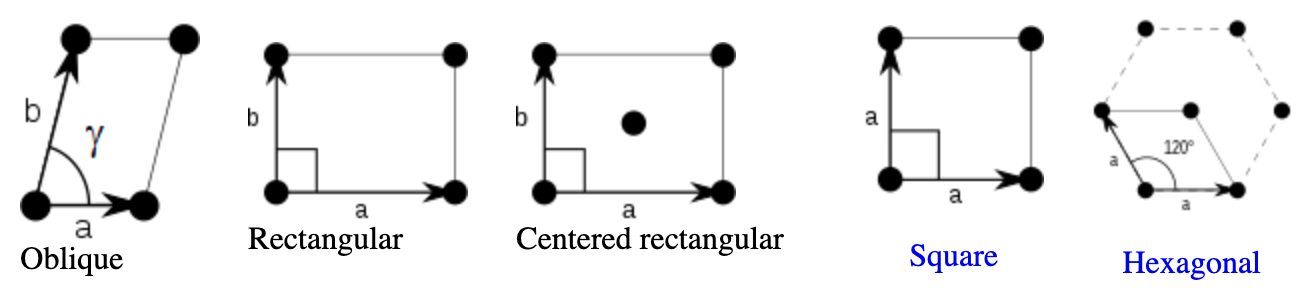}
		\caption{\small 2D periodic lattice is defined by two primitive vectors.  }	
	\label{fig:B}
\end{figure}

In 2D, a periodic lattice is defined by two primitive vectors $(\Delta \vx_1, \Delta \vx_2)$, and there are 5 different types of periodic lattices as shown in Figure \ref{fig:B}. If $\|\Delta \vx_1\| = \|\Delta \vx_2\|$, then the two possible lattices are square lattice and hexagon lattice. 

For Fourier analysis, we need to find the primitive vectors in the reciprocal space, $(\va_1, \va_2)$, via the relation: $\langle \va_i, \Delta \vx_j\rangle = 2\pi \delta_{ij}$, where $\delta_{ij} = 1$ if $i = j$, and $\delta_{ij} = 0$ otherwise. 

For a 2D periodic function $f(\vx)$ on a lattice whose primitive vectors are $(\va_1, \va_2)$ in the reciprocal space, define $\omega_k = k_1 \va_1 + k_2 \va_2$, where $k = (k_1, k_2)$ are a pair of integers (positive, negative, and zero), the Fourier expansion is $f(\vx) = \sum_k \hat{f}(\omega_k) e^{i \langle \omega_k, \vx\rangle}. $

See \url{http://lampx.tugraz.at/~hadley/ss1/crystaldiffraction/fourier/2dBravais.php} for more details. Figure \ref{fig:B} as well as Figure \ref{fig:group}(c) and (d) are taken from the above webpage. 

\subsection{More experiment results}

\subsubsection{Learned patterns}
\label{sec:pattern}
\begin{figure*}[h]
\centering
  \includegraphics[width=.9\linewidth]{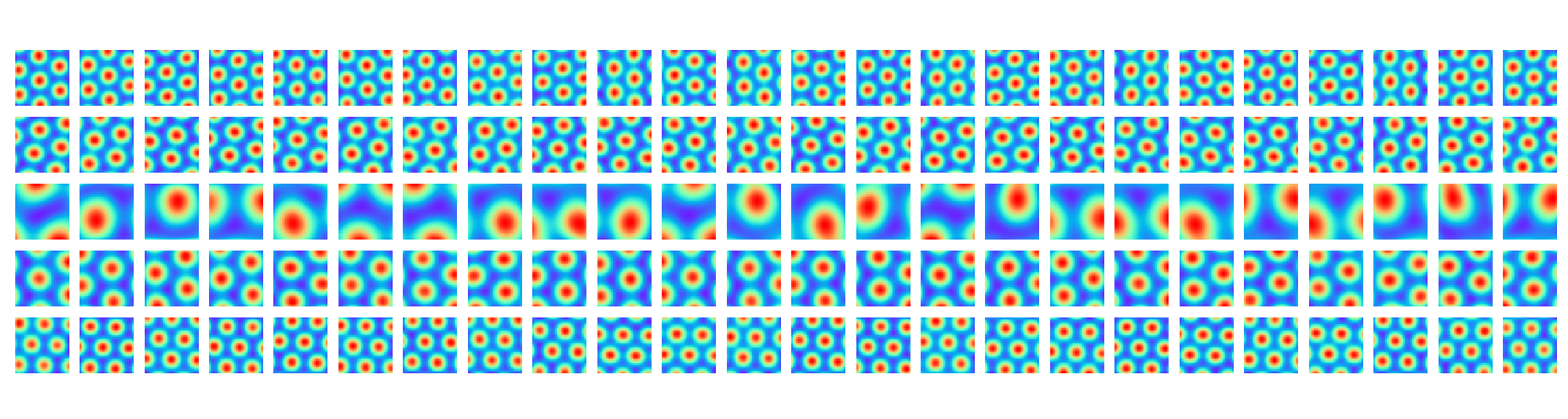} \\
  \caption{\small In the linear model, hexagon grid firing patterns are observed in the learned \( \vv(\vx) \). Each row displays the firing patterns of all the cells within a single module, with each module comprising 24 cells. The units illustrate the neuron activity throughout the entire 2D square environment. The figure presents patterns from five randomly chosen modules.}
  \label{fig: hexagon_linear}
\end{figure*}

\begin{figure*}[ht]
   \centering
   \begin{minipage}[b]{.45\textwidth}
  \centering         
   \includegraphics[width=.9\textwidth]{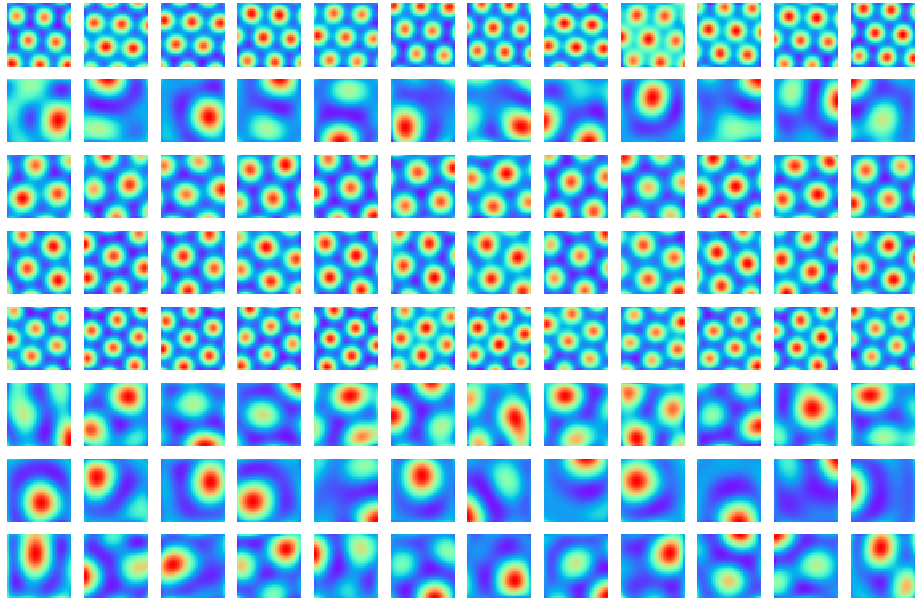}
   \\
   {\small (a) Tanh activation}
   \end{minipage}
   \begin{minipage}[b]{.45\textwidth}
  \centering
  \includegraphics[width=.9\textwidth]{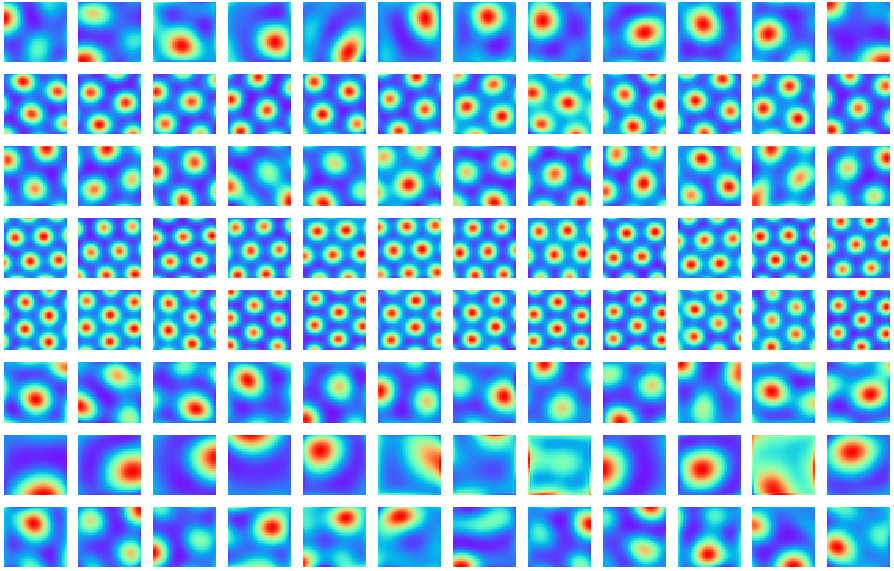}
    \\
   {\small (b) GeLU activation}
   \end{minipage}
   \caption{\small Results of the non-linear models. We randomly chose 8 modules and showed the firing patterns with different rectification functions. }
   \label{fig: hexagon_nonlinear}
\end{figure*}

In Figures \ref{fig: hexagon_linear} and \ref{fig: hexagon_nonlinear}, we show the learned grid patterns from the linear and non-linear models. 

\begin{figure*}[ht]
\centering
  \includegraphics[width=.9\linewidth]{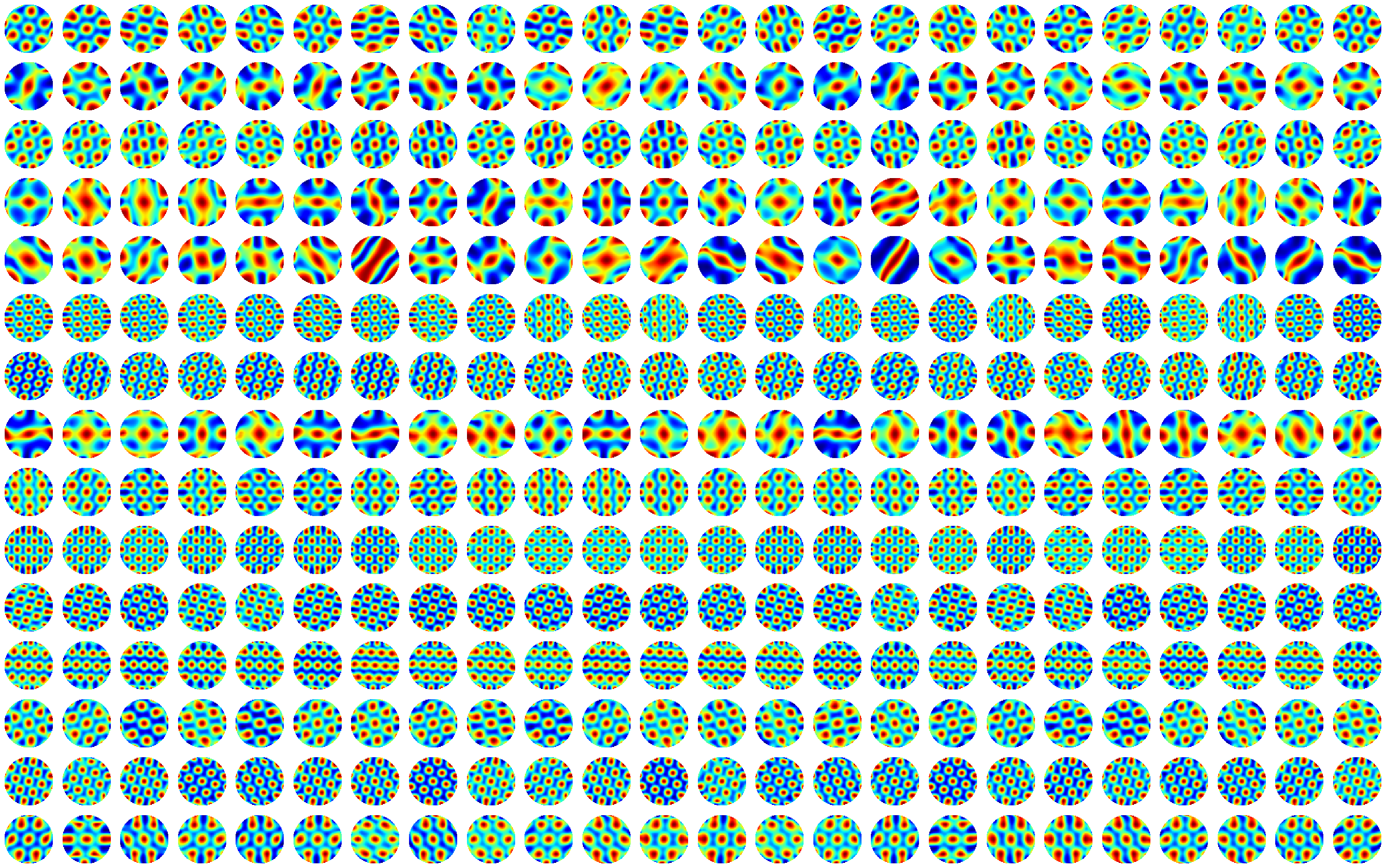} \\
  \caption{\small Autocorrelograms of the learned patterns for the linear model. }
  \label{fig: autocorr_linear}
\end{figure*}

\begin{figure*}[ht]
\centering
  \includegraphics[width=.9\linewidth]{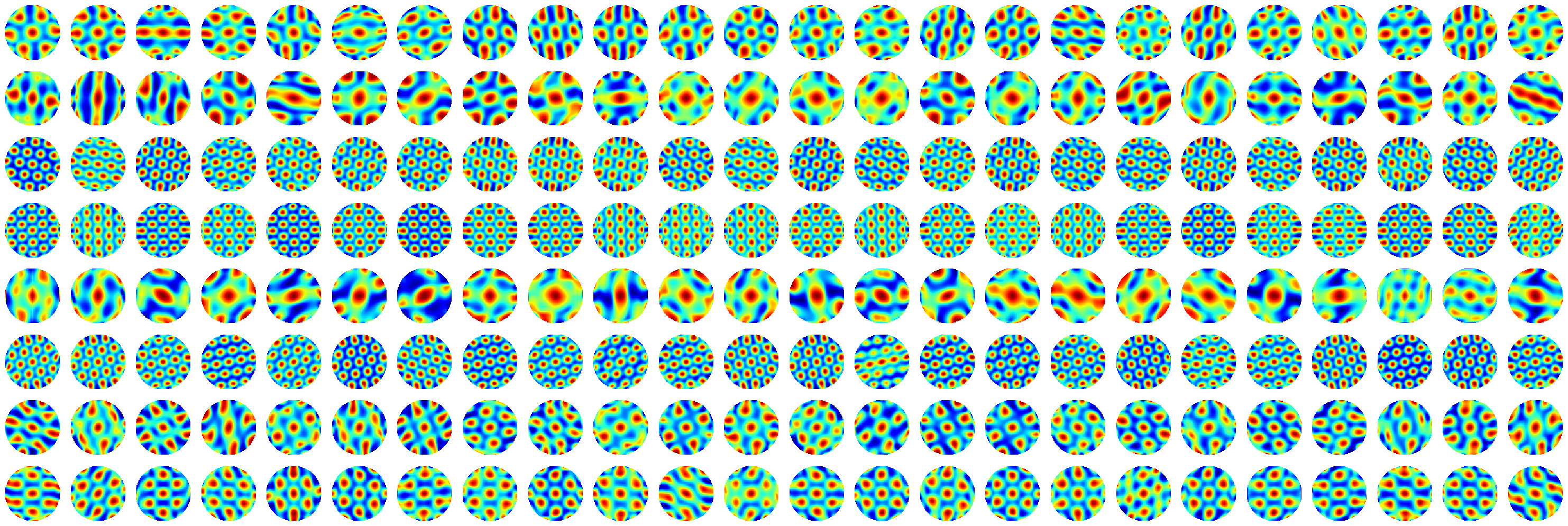} \\
  \caption{\small Autocorrelograms of the learned patterns for the non-linear model. }
  \label{fig: autocorr_nonlinear}
\end{figure*}

In Figures \ref{fig: autocorr_linear} and \ref{fig: autocorr_nonlinear}, we show the autocorrelograms of the learned grid patterns from the linear and non-linear models. 

\begin{figure}[ht]
   \centering
   \begin{minipage}[b]{.5\textwidth}
  \centering         
   \includegraphics[width=.99\textwidth]{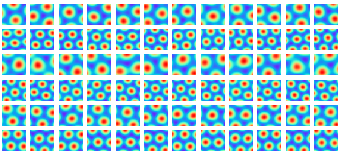}
   \\
   {\small (a) Block size is 12. }
   \end{minipage}
   \begin{minipage}[b]{.44\textwidth}
  \centering
  \includegraphics[width=.99\textwidth]{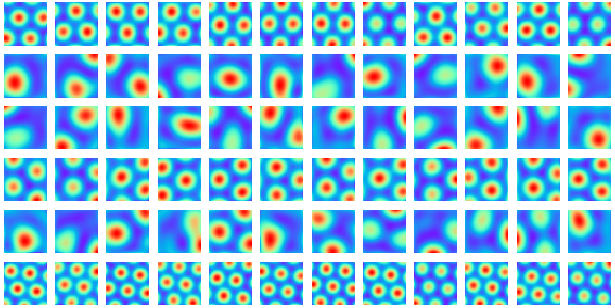}
    \\
   {\small (b) Block size is 36. }
   \end{minipage}
   \caption{\small For the linear model, the learned patterns of $\vv(\vx)$ with 12 and 36 cells in each block. We randomly select 6 blocks for each model and show 12 cells of those blocks. }
   \label{fig: different_size}
\end{figure}

We further tried with varying module sizes. Figure \ref{fig: different_size} visualizes the learned patterns when we fix the total number of grid cells but adjust the module size to $12$ or $36$. Importantly, the number or size of blocks doesn't impact the emergence of the hexagonal grid firing patterns.

\begin{figure*}[ht]
\centering
  \includegraphics[width=.9\linewidth]{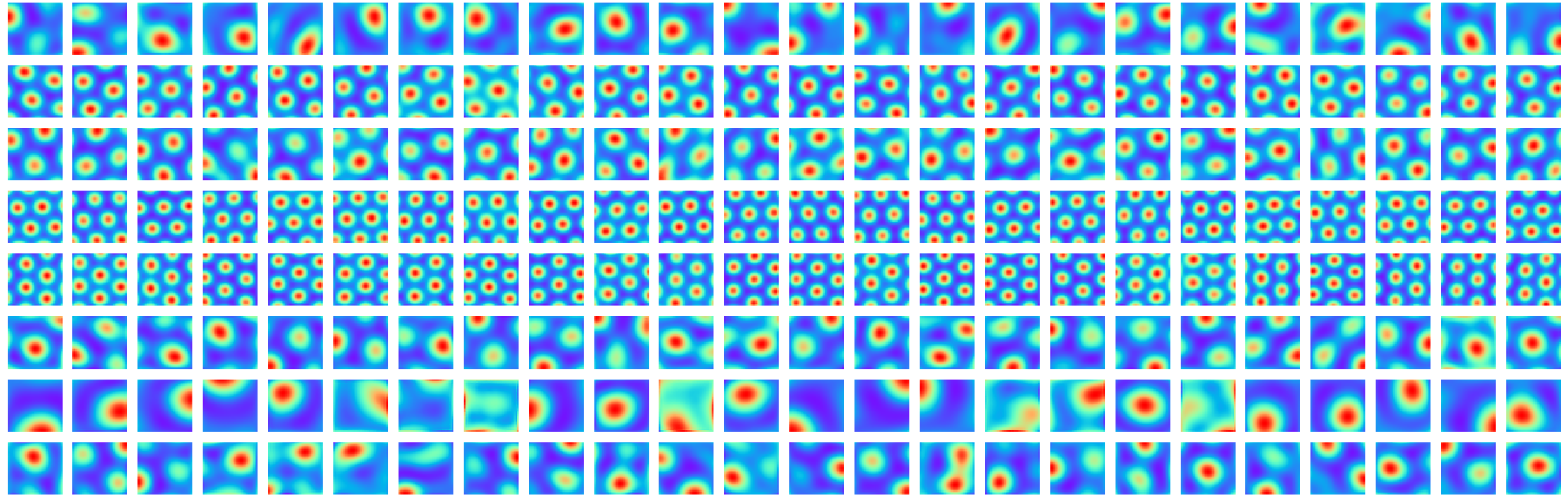} \\
  \caption{\small Firing patterns of the non-linear model with GeLU activation. }
  \label{fig: gelu}
\end{figure*}

Furthermore, for the non-linear model, we experimented with different rectification functions, including GeLU. Our evaluations of the learned patterns yielded a gridness score of $0.87$ and a ratio of grid cells at $78.65\%$. As depicted in Figure \ref{fig: gelu}, hexagonal grid firing patterns can emerge using diverse activation functions.

\begin{figure}[ht]
   \centering
   \begin{minipage}[b]{.48\textwidth}
  \centering         
   \includegraphics[width=.99\textwidth]{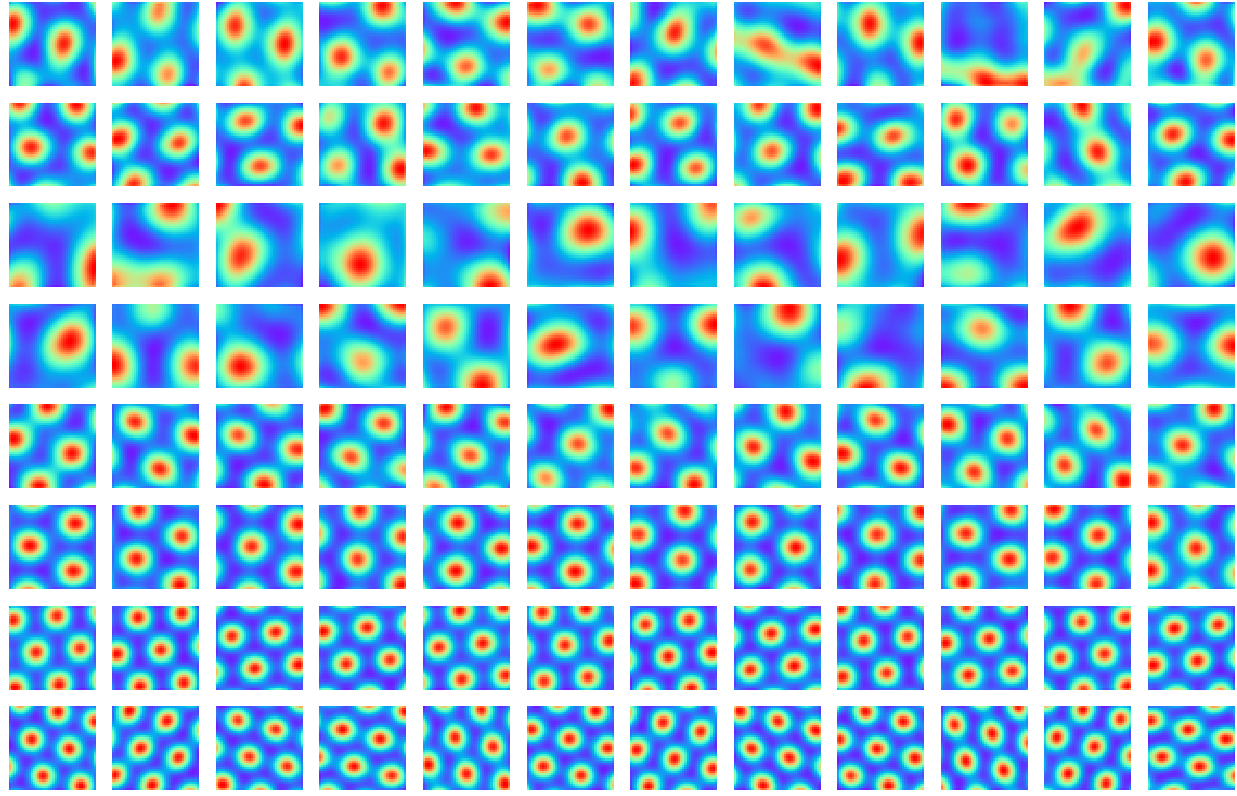}
   \\
   \end{minipage}
   \caption{\small Learned patterns with learnable scaling factor $s$. }
   \label{fig: learnable_s}
\end{figure}

Finally, for scaling factor $s$, we tried to learn it as a free parameter. In Figure \ref{fig: learnable_s}, we show the learned hexagonal patterns for the linear model with $12$ block size, which indicates that multi-scale grid patterns can be learned with or without learnable $s$. 

\subsubsection{Ablation studies}
In this section, we show ablation results to investigate the empirical significance of certain components in our model for the emergence of hexagon grid patterns. First, the emergence of hexagon patterns is dependent on conformal normalization for both linear and non-linear models. 

\label{sec:ablation}
\begin{figure}[ht]
   \centering
   \begin{minipage}[b]{.31\textwidth}
  \centering
  \includegraphics[width=.9\textwidth]{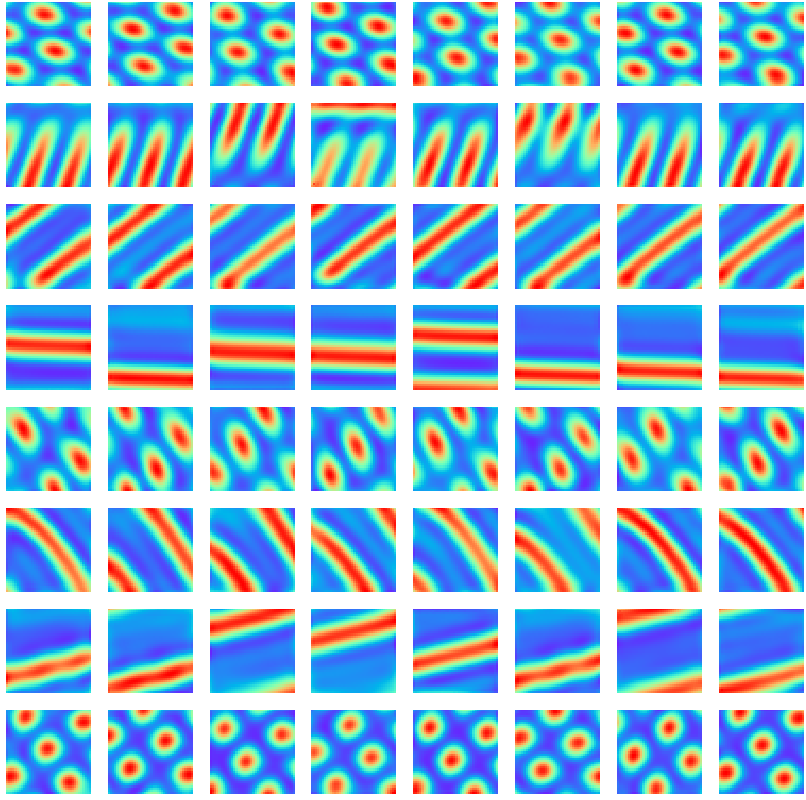}
    \\
   {\small (a) Linear model without conformal normalization}
   \end{minipage}
   \begin{minipage}[b]{.32\textwidth}
  \centering         
   \includegraphics[width=.9\textwidth]{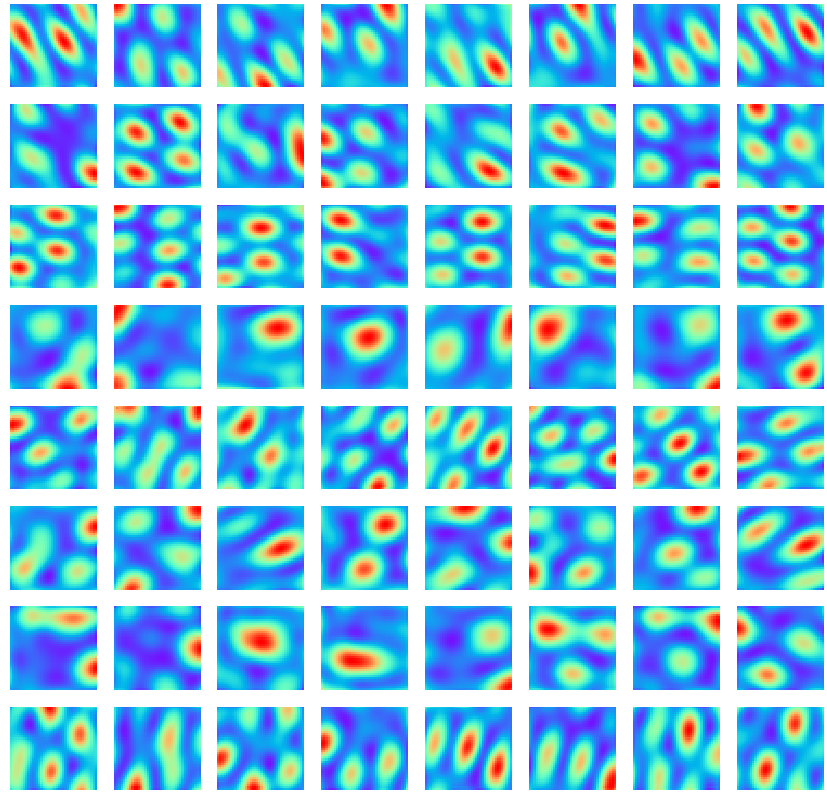}
   \\
   
   {\small (b) Non-linear model (Tanh activation) without conformal normalization}
   \end{minipage}
   \begin{minipage}[b]{.31\textwidth}
  \centering         
   \includegraphics[width=.9\textwidth]{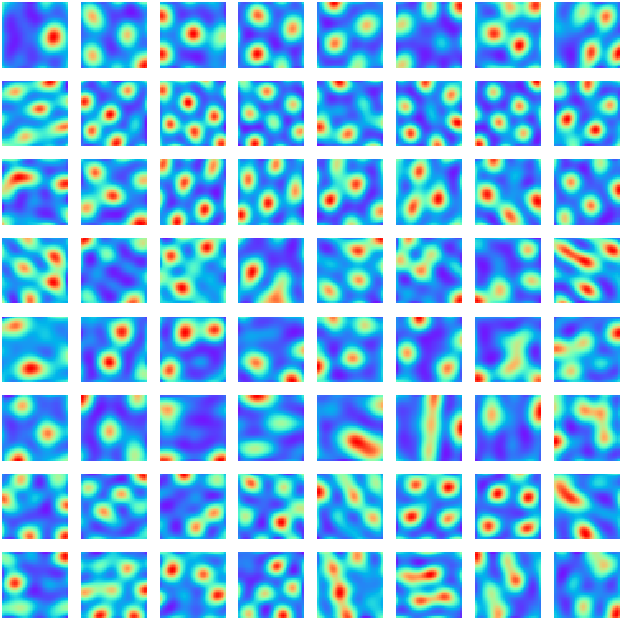}
   \\
   
   {\small (c) Non-linear model (GELU activation) without conformal normalization}
   \end{minipage}
   
   \caption{\small Results of ablation on certain components of the model. (a) Learned patterns without conformal normalization in the linear model. (b-c) Learned patterns without conformal normalization in the non-linear model. 
   }
   \label{fig: ablation}
\end{figure}

\end{document}